\documentclass[twocolumn,english,aps,showpacs]{revtex4}
\usepackage[T1]{fontenc}
\usepackage[latin9]{inputenc}
\usepackage{amsmath}
\usepackage{graphicx}
\usepackage{amssymb}

\makeatletter
\@ifundefined{textcolor}{}
{%
 \definecolor{BLACK}{gray}{0}
 \definecolor{WHITE}{gray}{1}
 \definecolor{RED}{rgb}{1,0,0}
 \definecolor{GREEN}{rgb}{0,1,0}
 \definecolor{BLUE}{rgb}{0,0,1}
 \definecolor{CYAN}{cmyk}{1,0,0,0}
 \definecolor{MAGENTA}{cmyk}{0,1,0,0}
 \definecolor{YELLOW}{cmyk}{0,0,1,0}
 }

\def\ket{\rangle}


\usepackage{babel}

\begin{document}

\title{Reply on Comments on "Observation of a Fast Evolution in a Parity-time-symmetric System"(Aixiv.1106.1550)}

\author{Chao Zheng$^{1}$, Liang Hao$^{1}$ and Gui Lu Long$^{1,2}$}

\address{$^{1}$State Key Laboratory of Low-dimensional Quantum Physics and
Department of Physics, Tsinghua University, Beijing 100084, P. R.
China\\
 $^{2}$Tsinghua National Laboratory for Information Science and
Technology,\\
 Beijing 100084, P. R. China }

\date{\today }
\begin{abstract}
Masillo \cite{masillo2} commented on our manuscript \cite{long}
"Observation of a Fast Evolution in a Parity-time-symmetric System",
pointing out a contradiction of our work with Ref. \cite{masillo}.
In this reply, we pointed out there is no disagreement between
Masillo's comment and our work in Ref. \cite{long}. The efficiency
cost pointed out in Ref.\cite{masillo} exists, namely to obtain the
PT-symmetric hamiltonian evolution, one has to make a measurement on
the auxiliary qubit and the auxiliary qubit is at state $|0\ket$
only probabilistically. This is reflected in the amplitude of the
spectrum in the NMR quantum simulation. As a result, we made a small
modification in a new version of the Ref. \cite{long}, and Fig. 2 of
Ref.\cite{long} has been replaced by spectra of two different
$\alpha$'s in order to illustrate this fact.
\end{abstract}


\maketitle



\end{document}